\documentclass[useAMS,usenatbib]{mn2e}
\usepackage{graphicx}
\usepackage{amssymb}
\usepackage{bm}
\usepackage{epstopdf}
\usepackage{color}

\title[Detecting hidden uncertainties in standard candles and clocks via improved hypothesis test]{Detecting hidden uncertainties in standard candles and clocks via improved hypothesis test}
\author[J. Wang and X. Meng]{
Jiaxin Wang$^{1,2}$
\thanks{E-mail:jxw@mail.nankai.edu.cn}
and
Xinhe Meng$^{1,2}$\\
$^{1}$Department of Physics, Nankai University, Tianjin 300071, China\\
$^{2}$State Key Laboratory of Theoretical Physics, Institute of Theoretical Physics,
Chinese Academy of Science, Beijing 100190, China}

\begin{document}
\date{}

\pagerange{\pageref{firstpage}--\pageref{lastpage}} \pubyear{2015}

\maketitle

\begin{abstract}
Tiny systematic uncertainty caused by cosmological hypotheses is hard to be detected, not only because the present observational errors are relatively large but also because hypothesis-induced uncertainty is indistinguishable from other sources of systematic errors. 
We introduce an efficient and sensitive method for detecting tiny systematic errors, which contain the cosmological-hypothesis-induced uncertainty and other secondary systematic errors, hidden behind residuals of chi-square analysis.
In this paper, we apply our analysis to JLA compilation of SN Ia observations and latest cosmic chronometer data-set. We find slight but noticeable evolutional feature in residuals of chi-square analysis under present systematic uncertainty control, when combining JLA samples with standard cosmological model.
Meanwhile, cosmic chronometer observation has no noticeable similar feature with various cosmological models, which may be covered up by relatively large observational uncertainties. 
Our method can be useful when various independent observational samples with high observational precision are available, since the cosmological hypothesis-induced error appears unbiasedly in all related data-sets.
\end{abstract}


\section{introduction}\label{sec1}
With deeper and more precise astrophysical observations coming around the corner, cosmologists are confident in discovering the nature of dark sections (ie., dark energy and dark matter) in the universe. More specifically speaking, future observations will enable us having more careful review upon various non-standard hypotheses and better estimation on theoretical parameters.

It is commonly accepted that the standard cosmological model $\Lambda$CDM although perfect enough to fit extremely well to most astrophysical observations, it contains unknown energy components in theoretical perspective. The key to the question about what the dark sections are and how to find imperfectness of standard cosmology, lies in seeking similar feature in deviation from ``standard'' prediction according to various observations.

Researchers are always concerning about removing all possible sources of significant systematic errors before cosmological parameter estimation with chi-square analysis which is critical for providing more robust and convincing results. We also notice that chi-square analysis is naturally not highly sensitive to some secondary systematic errors which are almost un-noticeable with respect to dominant observational uncertainties and are usually neglected in systematic error removing.

Since chi-square analysis relies on weighted summation of squares of residuals, the possible evolutional feature in the residuals, which represents the hidden systematic uncertainties, may be ignored in data analyzing.
This may result in tiny systematic errors, which are although far more less significant but also affecting the accuracy of estimation and blinding us from knowing the truth precisely. Since such small errors are seldomly concerned or detected in previous researches, we name it hidden uncertainties. Sources of hidden systematic uncertainties may come from unconcerned errors in observation, data selection, data analysis or theoretical hypotheses.

With present quality and quantity of observations, it is hard to find clear evidence for non-standard cosmological expansion history. This obstacle is caused by the fact that systematic uncertainty originated from possibly imperfect cosmological hypothesis, ie., constant dark energy term, is much smaller than dominant systematic errors in observed data samples. This means indications of new physics lie in the hidden uncertainties. Previous researches suggest that hypothesis-induced uncertainty may exist~\citep{herrera,zgb,zsn}.

Then the following task is about how to find the signal of hidden uncertainties against observational noises. Precisely speaking, the signal which we are looking for, performs as a specific form of systematic uncertainty while we estimating $\Lambda$CDM model parameters with observed samples. We separate the task into two steps, the first step is about introducing diagnostic method that suits various observations and theoretical predictions; and the second step concerns finding similar feature shared by various observations in diagnostic results.

The following section serves as a brief review about analysing schemes used in previous hypothesis tests. Then we move on to try detecting hidden systematic uncertainties with diagnostic function.

\section{hypothesis test}\label{sec2}

The basis of Bayes theorem lies in its explanation of probability which reads
\begin{equation}
P(A_i|B) = \frac{P(B|A_i)P(A_i)}{\Sigma_i P(B|A_i)P(A_i)},
\end{equation}
where $A$ can be regarded as a set of hypotheses with each specific model $A_i$, while $B$ represents observed data. $P(A_i|B)$ means the probability of acquiring model $A_i$ with given data-set $B$, which is usually called posterior probability; while $P(B|A_i)$, the likelihood function concerns about $\chi^2$ analysis
\begin{equation}
P(B|A_i) = N\cdot \exp\{-\frac{\chi^2(B|A_i)}{2}\}, \label{likelihood}
\end{equation}
represents the possibility of acquiring data-set $B$ with given model $A_i$. The priori, $P(A_i)$, does not affect too much about estimating parameters of models, thus is often given by experience.

The decision theory based on Bayesian theorem can not perform in full power in comparing different cosmological models, not only because the convincing form of priori can not be given precisely, but also due to the fact that the loss function, which is another critical part in making decision, is often neglected or given by experience. But without explicit forms of both priori and loss function (actually we just neglect their effects), we can still try to compare different hypotheses with Bayesian factor
\begin{equation}
K = \frac{P(B|A_i)}{P(B|A_j)}, \label{Bfactor}
\end{equation}
which in practice shares the same idea with Neyman-Pearson lemma.

Neyman-Pearson lemma defines that likelihood-ratio test rejects hypothesis $A_1$ in favour of $A_2$ when
\begin{equation}
\Lambda(B) = \frac{P(B|A_1,\theta)}{P(B|A_2,\theta)}\leq \eta, \label{N-P lemma}
\end{equation}
where the $P$ is the likelihood function we talked about in Eq.~(\ref{likelihood}), $\theta$ is a common parameter in both hypothesis, $\eta$ is a kind of threshold factor in discussing about significance.

Principally, we can choose the best theory only among all existing ones with Bayesian factor. But the consequent problem is we are not able to probe tiny systematic errors hidden behind observational errors, since chi-square analysis is not highly sensitive to tiny evolutional structure in residuals. This would make theoretical breakthrough hard to be found in phenomenological researches.

Hypothesis test that follows Ronald Fisher's principle seems to be a more direct approach, although the Fisher's approach had been accommodated in Neyman-Pearson test scheme~\citep{book}. But we would like to conduct null hypothesis test as a classical significance test in Fisher's scheme.

In classical significance test, we do not talk about the possibility that the null hypothesis can be proved, even if it is a truth~\citep{book}. The probability that a null hypothesis is not true can be expressed by the significance level of test. The classical significance test does not have to be accompanied by alternative hypotheses, which is the most different part between Fisher's approach and Neyman-Pearson lemma.

Imagine function $X$ is only related to observation, the value of it lies in $W$ where only part of $W$ can be predicted from a null hypothesis $H_n$. We define the value of $X$ predicted from null hypothesis lies in $w$, thus the probability that null hypothesis $H_n$ is not true reads
\begin{equation}
P(X\in W-w|H_n)=\alpha, \label{fisher}
\end{equation}
where $\alpha$ is the level of significance about falsification of null hypothesis $H_n$.

We can benefit from null hypothesis test as it offers a standard diagnostic criteria for conveniently assessing various models, since every model is presumed to be rejected equally at first.

Classical hypothesis test, or more specifically speaking in this paper, the null hypothesis test often requires precise value of parameters from hypothesis in order to perform in full power. So the problem is, the standard cosmological model does not predict precise value of most of the parameters, such as dimensionless Hubble constant $h$ and energy density of dust matter $\Omega_mh^2$. Considering about that, previous researchers focused on null hypotheses like ``The constant parameters are not dynamical.'', in order to use hypothesis test method in testing standard cosmology. Those null hypotheses are all logically equaling to standard cosmological model, only different in their expressions of diagnostic functions. The diagnostic function is actually acting as $X$ in Eq.~(\ref{fisher}). While focusing our discussion with flat space and low redshift observations, the most popular diagnostic function reads
\begin{equation}
D = \frac{H(z_i)^2-H(z_j)^2}{(1+z_i)^3-(1+z_j)^3}, \label{Dij}
\end{equation}
where $H(z)$ represents Hubble parameter at observed redshifts $z_i$ and $z_j$. This function is also named as two-point diagnostic function~\citep{shafieloo}, since it associates any two points in observation data-set at different redshifts.

The probability that the standard cosmological model is not true reads
\begin{eqnarray}
P(D\neq const.|\Lambda CDM) = 1,\\
P(D= const.|\Lambda CDM) = 0,
\end{eqnarray}
where $D$ represents a diagnostic function. This is an ideal but not practical rule~\citep{sahni}, which says the standard cosmological model is falsified if the values of diagnostic function deviate from one constant at any redshift.

According to the fact that realistic observations are coming with observational errors, a modified rule~\citep{clarkson} has been introduced which says the probability of falsification reads
\begin{equation}
P(D\neq const.|\Lambda CDM) = \frac{Len(B;D\neq const.)}{Len(B)}P(D), \label{modrule}
\end{equation}
where $Len(B)$ means the whole redshift range of observation data-set $B$, while $Len(B; D\neq const.)$ means the redshift range where diagnostic function deviates from a constant. Since the diagnostic function are reconstructed from data with observational error, its value at each specific redshift is expressed by a random variable with upper and lower limits. The limits are characterized by a level of confidence, for example, when $P(D) = 0.68$, the limits are chosen as $1\sigma$ deviation limits from the mean value. Thus the modified rule can be more applicable and reasonable.

\section{autocorrelation analysis}\label{sec3}

We find out that modified rule is still vulnerable to uncertainties that occurred in observations; besides, the signal of deviation between hypothesis and observation, even if can be observed with above rules, can not be clearly explained or quantified. Imagine we have observed some level of deviation according to rule Eq.~(\ref{modrule}), we still can not conclude that the deviation is due to the fact that our tested theory has some kind of level of deviation from the truth, since the deviation can be an illusion caused by systematic and/or random uncertainties in observation and data analysis.

There exists a more critical question for null test in practice: Are we expecting to find strong evidence of ruling out some model with null test while same level of evidence has not been spotted via chi-square analysis? We do not attempt to address this question in this article but we keep in mind that null hypothesis significance test may be assigned to a slightly different task discussed below.

In order to improve the null hypothesis test, we introduce autocorrelated analysis. The basic idea of our analysis is simple: Deviation between a theoretical hypothesis and truth can be manifested by observational inference in evolutional behavior of a constant parameter, where observational deviation against theoretical prediction acts like systematic error which unbiasedly appears in every related observational samples.

For revealing that particular systematic error we define an autocorrelation function $C(\theta_z)$ of diagnostic function as
\begin{equation}
C(\theta_z) = \langle I(\theta_z) \rangle_B, \label{average}
\end{equation}
where $\theta_z = (z_k-z_l)$, $\langle~\rangle_B$ represents average process in data-set $B$ with respect to $\theta_z$. The function $I$ is defined as
\begin{equation}
I(\theta_z) = D(z_k)-D(z_l), \label{I}
\end{equation}
where $D(z_k)$ represents available values of diagnostic which are related to observed quantity-set $\bold{q}$ at redshift $z_k$. The value of $D_z$ can be expressed as
\begin{equation}
D_z(\bold{q}_z) = D_{proj} + \delta_{r} + \delta_{s},
\end{equation}
where $D_{proj}$ means {\it true} value of $D$ that should be projected by precise observation without {\it any} error. For the observation suffers from systematic error and random error, their effects can be added to the projected value as $\delta_r$ and $\delta_s$.

In practice we assume the probability distribution of observed parameter $\bold{q}_z$ is gaussian, for example the possibility of finding the $i$th component $q_{z,i}$ in $\bold{q}_z$ at redshift $z$ equals to $p$ reads,
\begin{equation}
P(q_{z,i} = p) = P_0\exp \{-\frac{(p-\bar{q}_{i,z})^2}{\sigma_{i,z}^2} \} ,
\end{equation}
where $\{\bar{q}_{i,z}, \sigma_{i,z}\}$ is a set of observational output of quantity $q_i$ at redshift $z$. 

Although we are not able to express the probability distribution function of $D_z$ in a specific form, we can estimate its statistical expectation
\begin{equation}
\mathcal{E}[D_z] = D_{proj} + \mathcal{E}[\delta_r] + \mathcal{E}[\delta_s],
\end{equation}
where we presume $D_{proj}$ should depend on redshift when the null hypothesis is not true. Random error in real observation also depends on redshift, ie., random observational error is expected to grow when redshift goes larger, but we can average observational results at roughly the same redshift for minimizing influence of random error. 

We can also express the probability density function $g(D)$ of $D_z$ as
\begin{equation}
\mathcal{E}[D_z] = \int D g(D) dD = \int D(f_{proj} + f_r + f_s) dD,
\end{equation}
where $f_r$ and $f_s$ are related to $z$, while $f_{proj}$ should be a delta function independent of redshift only under null hypothesis.

Eq.~(\ref{average}) can be written as conditional expectation
\begin{equation}
C(\theta_z) = \mathcal{E}_Y[D_{z_k} - D_{z_l}],
\end{equation}
where $Y$ means $(z_k - z_l) = \theta_z$, as the condition of averaging. We can regroup the terms in $C(\theta_z)$ according to their dependency on $\theta_z$ like
\begin{equation}
C(\theta_z) = A + B(\theta_z),
\end{equation}
which means $A$ contains redshift independent terms. 
It is easy to calculate that the random error term reads
\begin{equation}
\int_Y D(f_{r,k}-f_{r,l})dD,
\end{equation}
where the offset character of random error should result in zero value of this term since theoretically we have $\int_Y D(f_{r,k}) dD=0$.
$A$ also contains
\begin{equation}
D_{p,k} - D_{p,l},
\end{equation}
under null hypothesis ($D_{p,k}$ represents value of $D_{proj}$ of $\mathcal{E}[D_{z_k}]$), but it is a $\theta_z$ dependent variable if null hypothesis is violated and thus acts like a (possibly secondary) systematic error as we mentioned. This hypothesis-induced systematic error is difficult to be separated from systematic uncertainties induced by other sources unless they have been carefully evaluated and corrected. 

But we can tell the difference between cosmological hypothesis-induced systematic uncertainty and other systematic uncertainties when apply same diagnostic quantity to different observations, since only cosmological hypothesis-induced systematic error is shared by every data-set.

The offset between random error terms requires enough samples under $Y$ condition, while in practice we change that critical condition into $(z_k-z_l)= \theta_z \pm \delta_\theta$ which means the average operations are taken in bins. Notice that for a sample-set with finite redshift range, the change in $\delta_\theta$ shall also change $\theta_z$ accordingly, this can be observed in Fig.~\ref{fig:Cbins}.

Different from previous treatments, we do not apply this scheme directly for detecting whether the $\Lambda$CDM is inconsistent with current observations. In the next section, we assign it with a more tangible mission: detecting secondary systematic error after chi-square analysis.

\section{tiny bias after fitting}\label{sec4}
Here we prove the autocorrelation analysis can serve as an efficient diagnostic method for detecting tiny bias after fitting cosmological parameters with chi-square analysis. We take $w$CDM as a general cosmological model which at low redshift reads,
\begin{equation}
h(z) = h\sqrt{\Omega_m(1+z)^3+(1-\Omega_m)(1+z)^{3\delta}},
\end{equation}
where $h(z) = H(z)/(100~km\cdot s^{-1}Mpc^{-1})$, and $\delta = 1+w$. $w$ represents the equation-of-state parameter of dark energy. The standard $\Lambda$CDM model can be recovered when $w = -1$. We presume the true Universe follows $w$CDM model with parameter $\{\Omega_m^*,h^*,\delta^* \}$. Notice that this assumption is a general setting for convenience which will not affect our proof.

In practice we build diagnostic function which focuses on Hubble constant. In the simplest case where the Hubble parameters at various redshift can be measured, the corresponding diagnostic function reads
\begin{equation}
D(z) = -5\log_{10}{[ \frac{h_{obs}(z)}{\sqrt{\Omega_m(1+z)^3+(1-\Omega_m)}}]}, \label{biasdiag}
\end{equation}
which contains observed quantities along with fiducial parameter $\Omega_m$. This is a little different from what we confronted in section~\ref{sec3} where diagnostic function only contains observed quantities. In practice it is hard to avoid introducing extra fiducial parameters, especially when using more precise observations, ie., SN Ia observation.

So generally we re-express the diagnostic function as $D(\bold{q},\gamma)$, where $\gamma$ means extra parameters. In Eq.~\ref{biasdiag}, we see $\gamma = \Omega_m$. The correlation spectrum can be expressed as
\begin{equation}
C(\theta_z) =\mathcal{E}_Y[ 5\log_{10}(\frac{h_{obs}(z_k)}{h_{obs}(z_l)}) - 5\log_{10}(\frac{E(z_k;\gamma)}{E(z_l;\gamma)}) ],
\end{equation}
where $E(z;\gamma) = \sqrt{\Omega_m(1+z)^3+1-\Omega_m}$. For an unbiased data fitting with $\Lambda$CDM model, the true Universe $h^*(z)$ must be characterized by $\{\Omega^*_m=\Omega^{f}_m, h^*=h^f, \delta^*=0 \}$. We set $\gamma$ with the best-fit value $\{\Omega^f_m,h^f\}$ from chi-square analysis, then the autocorrelation function $C(\theta_z)$ must yield zero values at each $\theta_z$, on the condition that every source of bias has been clearly corrected before and during chi-square analysis and we had presumed a right cosmological model.

Deviation from $C(\theta_z) = 0$ may be result from bias before and during chi-square analysis, and may also come from cosmological model we assumed if it deviates from the truth. Random uncertainties may also affect the diagnostic result, but with sufficient samples in each averaged bin, we can eliminate that influence and the robustness of the result against random error can be evaluated by using different width of average bin.

The essence of this analysis is finding evolutional structure in residuals, with respect to best-fitted prediction of chi-square analysis after dominant systematic uncertainties being corrected.

\section{confronting observation}\label{sec5}
In this section, we apply the correlation analysis to type Ia supernova (SN Ia) and cosmic chronometer (CC) observations, which are useful and carefully analysed low redshift observation in testing standard ($\Lambda$CDM) cosmological model.

\subsection{diagnostic function for SN Ia}
In order to use SN Ia data, we start with the relation between (dimensionless) luminosity distance and redshift
\begin{equation}
d_L(z) = (1+z)\int^z_0 \frac{dz'}{E(z')}, \label{distance}
\end{equation}
where $E(z')$, defined by dividing Hubble parameter $H(z)$ with Hubble constant $H_0$, is the dimensionless Hubble parameter. Practically, the observed information is represented by distance module $\mu$ which reads
\begin{equation}
\mu(z) = 5\log_{10}(d_L/h) + 5\log_{10}(3\cdot 10^8) , \label{distancemodule}
\end{equation}
where $h$ is the dimensionless Hubble constant, defined as $h = H_0/(100 km\cdot s^{-1}\cdot Mpc^{-1})$. For simplicity and accuracy, we define $\mu_0 = 5\log_{10}(3\cdot 10^8)$.

In terms of dimensionless distance, we have ignored the radiation and curvature terms in standard cosmological model since their contributions at such low redshift are negligible. So we have
\begin{equation}
E(z) = [\Omega_m(1+z)^3 + (1-\Omega_m)]^{1/2}.
\end{equation}

Through the observational perspective, distance module $\mu$ can be obtained through fitted light-curve parameters with SALT-II light-curve model~\citep{M14} of SN Ia observation, which reads
\begin{equation}
\mu_{obs} = m_B - (M_B^I + \Delta M_B^{I}) + \alpha X_1 -\beta C,
\end{equation}
where $m$ represents apparent peak magnitude of luminosity at B-band with $M^{I}$ as its corresponding intrinsic magnitude of luminosity; $\Delta M_B^{I}$ means stellar mass related correction which is zero when $M_{stellar}< 10^{10}M_\odot$; $X_1$ and $C$ are stretch parameter and color parameter respectively, which describe the properties light-curve with $m$ together. $\alpha$ and $\beta$ are their constant coefficients.

Considering both theoretical and observational expression of distance module, the diagnostic function is designed as
\begin{equation}
D = (m_B - \Delta M^{I}_B + \alpha X_1 -\beta C) - 5\log_{10}{d_{L,fit}}.
\end{equation}
The diagnostic quantity of this function is $M^I_B-5\log_{10}h$, where the intrinsic magnitude of supernova has also been assumed to be a constant. The distance term $d_{L,fit}$ represents luminosity distance described by Eq.~(\ref{distance}) obtained with best-fitted $\Omega_m$ value.

\subsection{test with JLA compilation}
The JLA (Joint Light-curve Analysis, which is jointly conducted by SDSS and SNLS collaborations) compilation~\citep{JLA} is adopted as our SN Ia samples. The systematic uncertainties from SALT-II light-curve analysis have been thoroughly evaluated with calculation and simulation in previous researches~\citep{G10,M14} and careful systematic uncertainty analysis had been adopted in original report~\citep{JLA}. The best-fit values of cosmological parameters are constrained after bias corrections in light-curve parameters included in the released data which is adopted in our analysis.

For evaluating values of diagnostic function at various redshifts, we adopt the JLA's original report where $\alpha=0.141$, $\beta=3.101$, $\Delta M_B^I = -0.07$ which are best-fit parameter values with systematic uncertainties and bias corrections included in chi-square analysis. The joint analysis also conduct parameter constraints according to joint analysis of Planck2013~\citep{plk2013} and JLA, which yield slightly higher matter density $\Omega_m = 0.305\pm 0.010$ than from JLA alone which gives $\Omega_m = 0.295\pm0.034$ while other parameters suffers smaller differences. Avoiding from bringing possibly extra unknown tiny uncertainty from combining two independent data-sets~\citep{plk2015}, we adopt best-fitted $\Omega_m = 0.295$ according to JLA sample only.

We also generated a simple set of mock observational data which also contains 740 samples for comparing their results. The mock data-set is produced as procedure described bellow~\citep{karpenka}:

1.~The redshift is drawn independently from $z_i\sim U(0,1.3)$, where $U(a,b)$
denotes a uniform distribution in the range $[a,b]$.

2.~The predicted distance module at redshift $z_i$ $\mu_i(z_i,\Omega_m=0.3,h=0.7)$ is calculated using Eq.~(\ref{distancemodule}).

3.~The hidden variables $M_i$ (intrinsic magnitude), $X_{1,i}$ (stretch parameter) and $Ci$ (color parameter) are drawn from the respective distributions $\hat{M_i}\sim \mathcal{N}(-19.03,0.02)$, $X_{1,i}\sim U(-2.862,2.337)$ and $C_i\sim U(-0.250,0.260)$,
where $\mathcal{N}(\mu,\sigma^2)$ denotes a normal (Gaussian) distribution with mean $\mu$ and variance $\sigma^2$.

4.~The value of $m_{B,i}$ is calculated using the Phillips relation $m_{B,i}=\mu(z_i,0.3,0.7)+M_i-\alpha X_{1,i}+\beta C_i$.

5.~The simulated observational data are drawn independently from the distributions $\hat{z}_i\sim \mathcal{N}(z_i,\sigma_z^2)$, $\hat{m}_{B,i}\sim \mathcal{N}(m_{B,i},\sigma_{m,i}^2)$, $\hat{X}_{1,i}\sim \mathcal{N}(X_{1,i},\sigma_{X,i}^2)$ and $\hat{C}_i\sim \mathcal{N}(C_i,\sigma_{C,i}^2)$, where $\{\sigma_{m,i},\sigma_{X,i},\sigma_{C,i}\}$ are drawn randomly from uniform distributions $\sigma_{m,i}\sim U(0.09,0.175)$, $\sigma_{X,i}\sim U(0.018,1.641)$ and $\sigma_{C,i}\sim U(0.012,0.107)$. All upper and lower limits in uniform distributions mentioned in each step are obtained from real JLA data-set.

We do not re-constrain the fiducial cosmology which has $\{\Omega_m =0.3, h = 0.7,\alpha=0.12,\beta=3.2\}$ according to mock data, but reproduce the diagnostic function $C(\theta_z)$ with fiducial parameter-set $\{\Omega_m,\alpha,\beta\}$ since we do not doubt whether chi-square fitting scheme in real SN Ia analysis may introduce systematic bias.

\begin{figure}
\includegraphics[width=0.45\textwidth]{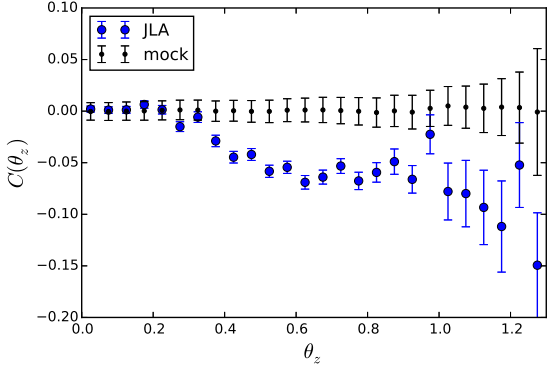}
\caption{Autocorrelation function constructed according to JLA compilation (with best-fit parameter setting) and the general mock data-set separately. The evaluation errors are at $1\sigma$ level, we can observe the results from real observation deviate at more than $2\sigma$ level from those of general data-set at redshift higher than about $0.3$ and lower than about $0.8$.}
\label{fig:C}
\end{figure}

The simulation process adopted here is far from realistic, but it is sufficient for comparing results from correlation analysis with simulated and real observed data-set. The point is that we are trying to estimate the influence from random error and finite sample number in blurring signals in our analysis. If the result from real data lies within the region of simulated result, then we conclude there exists no noticeable hidden systematic uncertainties. Thus our focus here is about evaluating blurring range (as shown by black points with error bars in Fig.~\ref{fig:C}) from random error rather than conducting extremely realistic simulations of observation.

Notice that in step $5$ we concern analyzing uncertainties in light-curve parameters, which may lead to un-expected systematic-like uncertainties in final result since observed sample number is finite. In order to include various possibilities, we generate a general mock data-set (displayed in Fig.~\ref{fig:C}) in which mean value of light-curve parameters are set by true values, and extend the error to the maximum $2\sigma$ level as $\sigma_{m,i} = 0.35$, $\sigma_{C,i}= 0.214$ and $\sigma_{X,i}=3.282$. Systematic-like fluctuations in each correlation spectrum from a single mock data-set can be well included by $2\sigma$ evaluation uncertainties of this general data-set as shown in Fig.~\ref{fig:C}.

The means and standard deviations of values of $C(\theta_z)$ is estimated by the following process: First, for each SN Ia sample at redshift, ie., $z_k$, we can estimate the distribution of value of diagnostic function and characterize it by its mean and standard deviation. Then for each pair of SN Ia samples, distribution of value of Eq.~(\ref{I}) can be estimated. We also characterize each $I$ with its mean and standard deviation. In order to realize average condition $Y$ as we previously mentioned, mean value and standard deviation of $C$ in each $\theta_z$-bin are estimated through unconstrained averaging with standard weighted least-squares method, which reads
\begin{equation}
C\pm \delta C = \frac{\Sigma_i w_i I_i}{\Sigma_i w_i} \pm(\Sigma_i w_i)^{-\frac{1}{2}},
\end{equation}
where the weight $w_i$ is
\begin{equation}
w_i = \frac{1}{(\delta I_i)^2}.
\end{equation}
$C$ and $\delta C$ stand for mean value and standard deviation of value of $C(\theta_z)$ in each bin, while $I_i$ represents each mean value in the same bin with its wight calculated from standard deviation. Notice that we estimate $D(z)$ in bins directly before calculating $C(\theta_z)$ through the same average method, which is displayed as bottom panel in Fig.~\ref{fig:Cvar}.

\begin{figure}
\includegraphics[width=0.45\textwidth]{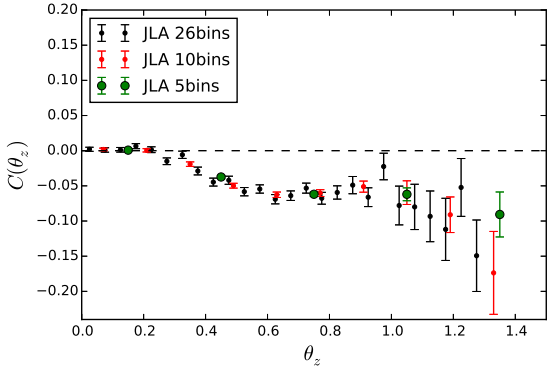}\\
\caption{The same function according to real data in Fig.~\ref{fig:C} but calculated with different widths of bins, showing the binned estimation of $C(\theta_z)$ is robust.}
\label{fig:Cbins}
\end{figure}

\begin{figure}
  \includegraphics[width=0.45\textwidth]{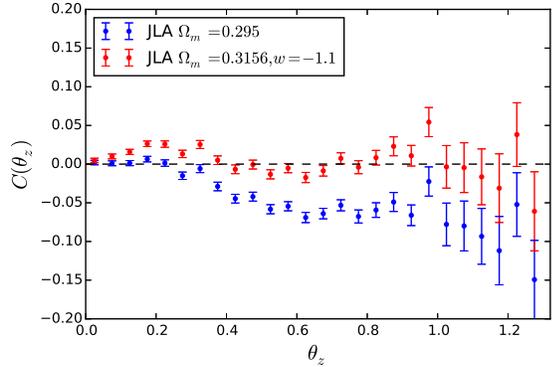}\\
  \includegraphics[width=0.45\textwidth]{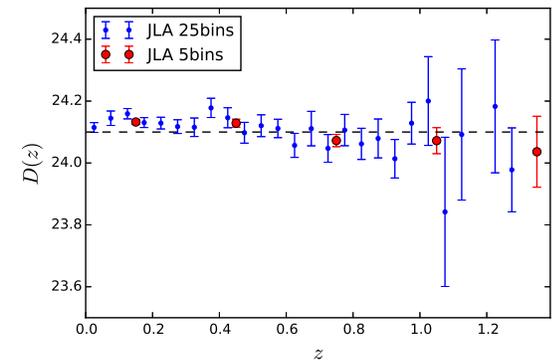}\\	
  \caption{Top: Autocorrelation function constructed according to JLA compilation with $\Lambda$CDM which has best-fit parameter setting (white), in contrast with spectrum (red) with $w$CDM which changes $w = -1.1$ and $\Omega_m= 0.3156$. Bottom: Estimating $D(z)$ in redshift bins, from which we can observe deviation clearly. This figure verifies the result of $C(\theta_z)$ in Fig.~\ref{fig:Cbins}.}
  \label{fig:Cvar}
\end{figure}

The autocorrelation spectrums from JLA and mock data-sets are displayed in Fig.~\ref{fig:C}, where we conduct expectation operation $E_Y[~]$ in bins which means $Y$ represents $|z_k -z_l| = \theta_z\pm 0.025$. The robustness of taking different binning widths is shown in Fig.~\ref{fig:Cbins}, we also tried randomly dropping one third number of samples in the full data-set but the result is negligibly affected.

Since the mock data-set is free from systematic uncertainties, its spectrum remains relatively horizontal as we expected. We also checked the final error in reconstructed value of $D$, the mock data can provide average final standard deviation $0.178$ which is exactly the same as that from real data.

Real observational result shows an noticeable deviation from horizontal line, we think that pattern results from either un-concerned tiny systematic uncertainties in data or cosmological hypothesis.

We also find that replacing standard cosmology by $w$CDM model with EoS parameter of dark energy lower than $-1$ may ease the decreasing pattern in $C(\theta_z)$, as shown Fig.~\ref{fig:Cvar} which indicates the $C(\theta_z)$ of real SN Ia observation is sensitive to cosmological model parameters. But we are not suggesting that $w$CDM with phantom-field can fit better with current covariance matrix provided. As a matter of fact, $w=-1.1$ only meets the edge of $1\sigma$ confidence range in the likelihood of given fitting result~\citep{JLA}. Fixing that deviation requires more exquisite analyzing processes and precise bias corrections in the future.

\begin{figure}
\includegraphics[width=0.45\textwidth]{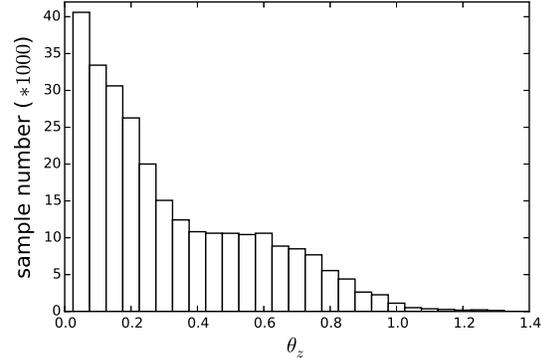}\\
\caption{Number of sapmles included in each bin (when calculated in $26$ bins), where one sample indicates a pair of two observational points at different redshift. For JLA compilation which contains 740 observed points, the total amount of ``sample'' is about 270 thousand.}
\label{fig:Ccount}
\end{figure}

\subsection{diagnostic function for cosmic chronometers}
Cosmic chronometers (CC) are relatively more direct probes for cosmic expansion rate $H(z)$, which relates to age difference of early type galaxies (ETGs) as
\begin{equation}
H(z) = -\frac{1}{1+z}\frac{dz}{dt},
\end{equation}
where $dt$ represents age difference between galaxies with redshift difference $dz$. The age differences were historically measured in two methods due to the fast development in astrophysical observations and analyses. With relative age based method, the age of ETGs are measured by stellar population synthesis models, while the $D4000_n$ method concerns the spectral $4000~\AA$ break feature in ETGs~\citep{M15} which is approximated to be linearly related to galaxy age as
\begin{equation}
D4000(Z) = A(SFH,Z/Z_0)\cdot age +B(Z),
\end{equation}
where $Z$ represents metallicity, $B$ and $A$ are constant parameters, SFH means the parameter $A$ also depends on star formation history. Cosmic expansion rate at redshift $z$ is defined as
\begin{equation}
H(z) = -\frac{1}{1+z}A\frac{dz}{dD4000},
\end{equation}
where $dD4000$ means the average difference in 4000 A break corresponding to redshift difference $dz$.

In order to make our analysis directly comparable with that to SN Ia, we point out a simple diagnostic function which shares similar diagnostic quantity $h$ as
\begin{equation}
D = -5\log_{10} \{ \frac{h(z)}{\sqrt{\Omega_m(1+z)^3+(1-\Omega_m)}} \},
\end{equation}
where $h(z) = H(z)$. We adopt the simplest model for low redshift expansion prediction as we did for SN Ia analysis under $\Lambda$CDM hypothesis.

\subsection{test with CC-M15}
The latest update (M15) of CC data-set was reported by M.~Moresco~\citep{M15} where two new measurements of Hubble parameter with D4000 method were added to the old CC compilation~\citep{CC1,CC2}. Two observed points at redshift $0.48$ and $0.88$ are removed, since the relative estimation errors in those $H(z)$ are too large to protect their sampling from touching negative value. We name the new compilation as CC-M15 (as shown in Tab.~\ref{tab0}), the systematic errors and consistence between different measurements were carefully analysed~\citep{M12a,M12b}. Since the quality of this measurement is still low in comparison with SN Ia, its capacity in distinguishing cosmological parameters is thus not very powerful. But M15 update greatly increased its redshift depth to about $2.0$, which makes CC the most deep cosmological data-set independent from CMB at present. We fit $\Lambda$CDM model with the data-set and get best-fit results: $\Omega_m=0.318$, $h=0.689$ with $\chi^2_{min}/dof = 0.719$.

\begin{table}
\caption{ CC-M15 compilation of H(z) measurements, we had dropped 2 samples at redshift $0.48$ and $0.88$ due to their large measurement errors.
}
\label{tab0}
\begin{tabular}{ccc}
\hline
\hline
z &  H(z)~($km/s/Mpc$) & $\sigma_{\rm H}$ \\
\hline
0.09&	69&	12\\
0.17&	83&	8\\
0.179&	75&	4\\
0.199&	75&	5\\
0.27	&77&	14\\
0.352&	83	&14\\
0.4&	95	&17\\
0.593&	104&	13\\
0.68	&92&	8\\
0.781&	105&	12\\
0.875&	125&	17\\
0.9&	117&	23\\
1.037&	154&	20\\
1.3	&168	&17\\
1.363	&160&	33.6\\
1.43&	177&	18\\
1.53&	140&	14\\
1.75&	202&	40\\
1.965	&186.5&	50.4\\
\hline
\hline
\end{tabular}
\end{table}

\begin{figure}
  \includegraphics[width=0.4\textwidth]{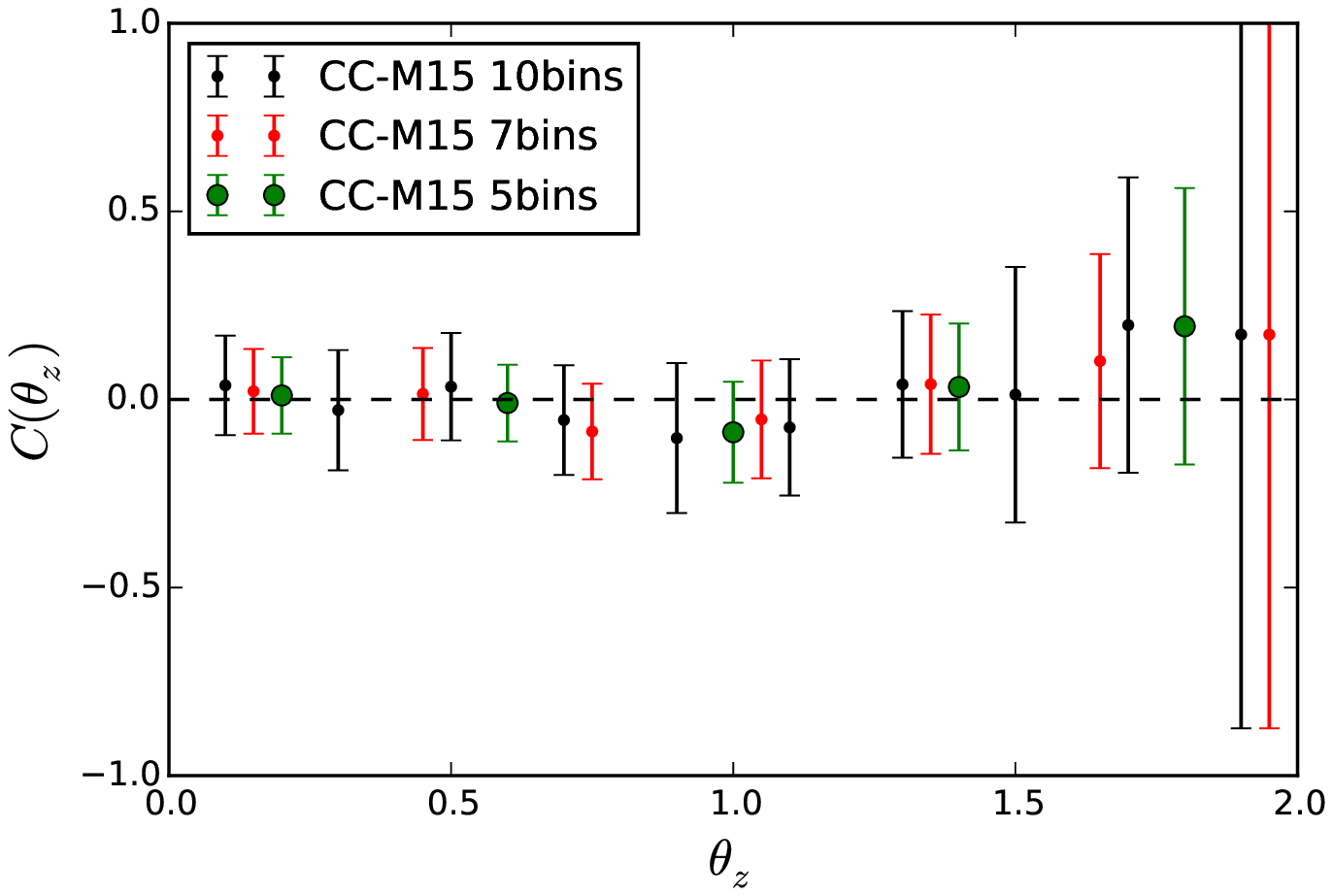}\\
  \includegraphics[width=0.4\textwidth]{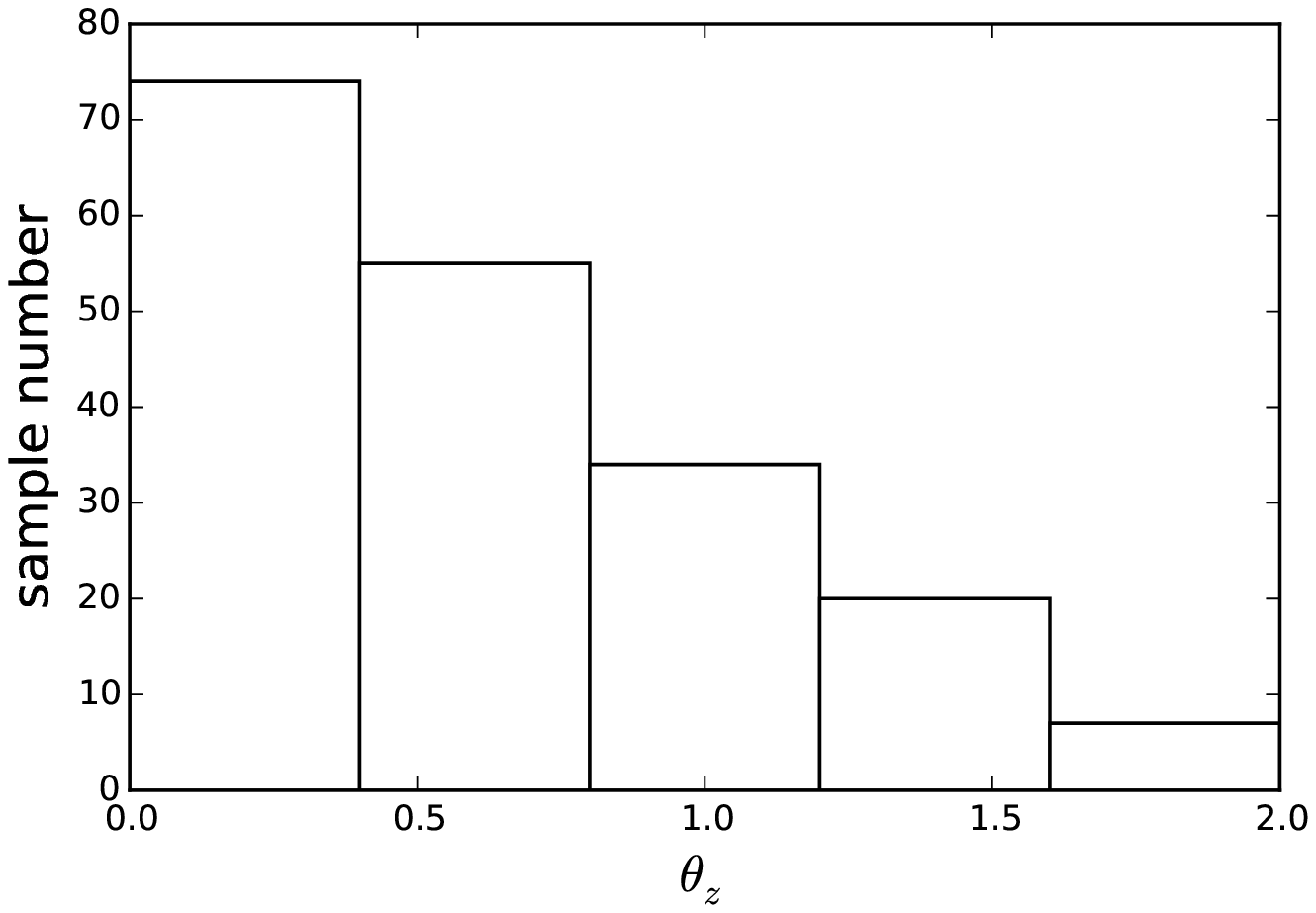}\\
  \caption{{\it Top}: Autocorrelation functions calculated according to CC-M15 H(z) data-set with different number of bins, showing result is also robust. {\it Bottom}: Number of samples contained in five bins, the definition of ``sample'' is the same as that in Fig.~\ref{fig:Ccount}.}
  \label{fig:Ch}
\end{figure}

The autocorrelation functions with various number of average bins are displayed in Fig.~\ref{fig:Ch}. there is no noticeable leftover systematic errors. Uncertainties in evaluating correlation spectrum binned values are large enough to cover the zero line.
We try to fit CC data with $w$CDM and $w_z$CDM~\citep{polarski,wzcdm} models and get Tab.~\ref{tab1}. Replacing $\Lambda$CDM by those two model with fitted values has negligible effect on the spectrum. This is mainly due to the shortcoming in sample number and large observational errors.

\begin{table}
\caption{Best fit values of $w$CDM and $w_z$CDM model with CC-M15.
}
\label{tab1}
\begin{tabular}{cccccc}
\hline
\hline
model & $\Omega_m$  & $h$ & $w_0$ & $w_a$ & $\chi^2_{min}/dof$ \\
$w$CDM& 0.311 & 0.681 & -0.913 & -- & 0.763\\
$w_z$CDM& 0.226& 0.693& -0.966& 0.900& 0.809\\
\hline
\hline
\end{tabular}
\end{table}

According to the analysis with CC data-set, we can not capture similar information as which from JLA data-set. No tiny systematic uncertainty can be measured. This means with present accuracy of CC data, constraining standard cosmological model and even non-standard models is not significantly affected by hidden systematic errors, although the results from data fitting also suffers from relatively low precision. But the physics behind CC observation is much more simpler and the data extracting process is affected by less systematic uncertainties, we consider that as the advantage of cosmic chronometers against SN Ia. More observational samples of cosmic chronometers are highly expected in order to conduct cross check with SN Ia data and realise the final goal of our test.

\section{discussion}\label{discussion}
In this paper we introduced a correlation scheme for detecting hidden systematic uncertainty after chi-square analysis. The goal of our work is about providing efficient and sensitive methodology for estimating possibly tiny unconcerned or cosmological hypothesis-induced uncertainties. With improvement in quality and quantity of observed samples in the future, this method can also serve as an useful tool for seeking indication of imperfectness in standard cosmological model.

With currently observed SN Ia data, we find existence of tiny systematic errors after previous analysing which suggests there may be unconcerned subdominant but still detectable systematic uncertainties or indication to non-standard cosmological model. Although such systematic deviation can be catched by our analysis, it has slight influence in likelihoods of estimated parameters. We suggests that addressing that issue can help in providing better constraints on cosmological predictions. According to the latest research~\citep{mag}, the intrinsic luminosity of SN Ia is affected by star formation status of its host galaxy, thus we can not exclude such possible selection bias in observation at present. We hope to overcome those difficulties in future research when better observational correction in large sample of SN Ia observations is available.

In terms of CC data, we find the correlation analysis is affected by the lack of sample number and quality. But current data seems fine with no noticeable leftover systematic uncertainties. We are looking forward to see more CC samples coming in the future, then comparing the correlation spectra from SN Ia and CC can tell whether there exists common feature, which is very likely induced by uncertainty from cosmological model.

Our result seems to be unexpected, since intuitively we don't think there still exist detectable systematic uncertainties after impressive bias control adopted by~\citet{JLA}. The truth is that systematic deviation in Fig.~\ref{fig:C} is not detectable at all in Hubble diagram, but only noticeable through autocorrelated analysis. Thus there is no contradiction between our result and that from JLA original analysis, the point is that we are looking into the evolutional structure of residuals in Hubble diagram. Similar to our method, fitting parameters in redshift bins may also address such issue, but the signal may also be blurred by observational uncertainties.

Although we only focus on low redshift astrophysical observations, the analyzing scheme can be applied to other measurements of theoretical parameters as well.

\section*{Acknowledgements}
We thank Prof.~Gang~Chen and Dr.~Z.~Zhang for providing precious suggestions during this work. We also thank Prof. D. Polarski for showing us useful references.


\newpage
\begin{thebibliography}{99}

\bibitem[\protect\citeauthoryear{Benitez-Herrera et al.}{2013}]{herrera}
Benitez-Herrera S., {\it et. al.}, MNRAS, 436, 854.

\bibitem[\protect\citeauthoryear{Betoule et al.}{2014}]{JLA}
Betoule M., et al., AA. 568 A22.

\bibitem[\protect\citeauthoryear{Chevallier \& Polarski}{2001}]{polarski}
Chevallier M., Polarski D., Int. J. Mod. Phys. D10, 213.

\bibitem[\protect\citeauthoryear{Guy et al.}{2010}]{G10}
Guy J., et al., AA. 523 A7.

\bibitem[\protect\citeauthoryear{Ichikawa \& Takahashi}{2008}]{ichikawa}
Ichikawa K., Takahashi T., JCAP 0804, 027.

\bibitem[\protect\citeauthoryear{James}{2006}]{book}
James F., ``{\it Statistical Methods in Experimental Physics}'' 2nd Edition, World Scientific Press, 2006.

\bibitem[\protect\citeauthoryear{Karpenka}{2015}]{karpenka}
Karpenka N. V., ISBN 978-91-7447-953-9, arXiv: 1503.03844.

\bibitem[\protect\citeauthoryear{Linder}{2003}]{wzcdm}
Linder E. V., PRL. 90, 091301.

\bibitem[\protect\citeauthoryear{Moresco et al.}{2012a}]{M12a}
Moresco M., et al., JCAP 08, 006.

\bibitem[\protect\citeauthoryear{Moresco et al.}{2012b}]{M12b}
Moresco M., et al., JCAP 07, 053.

\bibitem[\protect\citeauthoryear{Mosher et al.}{2014}]{M14}
Mosher J., et al., ApJ. 793 16.

\bibitem[\protect\citeauthoryear{Moresco}{2015}]{M15}
Moresco M., MNRAS, in press, arXiv: 1503.01116.
\bibitem[\protect\citeauthoryear{Nesseris \& Perivolaropoulos}{2005}]{nesseris}
Nesseris S., Perivolaropoulos L., PRD 72, 123519.
\bibitem[\protect\citeauthoryear{Planck Collab.}{2013}]{plk2013}
Planck Collab., arXiv:1303.5076.
\bibitem[\protect\citeauthoryear{Planck Collab.}{2015}]{plk2015}
Planck Collab., arXiv:1502.01589.

\bibitem[\protect\citeauthoryear{Rigault et al.}{2015}]{mag}
Rigault M., et al., ApJ. 802, 20.

\bibitem[\protect\citeauthoryear{Sahni, Shafieloo \& Starobinsky}{2008}]{sahni}
Sahni V., Shafieloo A., Starobinsky A., PRD 78, 103502.

\bibitem[\protect\citeauthoryear{Shafieloo et al.}{2012}]{shafieloo}
Shafieloo A., et al., PRD 86, 103527.
\bibitem[\protect\citeauthoryear{Seikel \& Clarkson}{2013}]{clarkson}
Seikel M., Clarkson C., arXiv:1311.6678.

\bibitem[\protect\citeauthoryear{Simon, Verde \& Jimenez}{2005}]{CC1}
Simon J., Verde L., Jimenez R., PRD 71, 123001.

\bibitem[\protect\citeauthoryear{Stern et al.}{2010}]{CC2}
Stern D., et al., JCAP 02, 008.

\bibitem[\protect\citeauthoryear{Yahya et al.}{2014}]{seikel}
Yahya S., et al., PRD 89, 023503.

\bibitem[\protect\citeauthoryear{Yu, Yuan \& Zhang}{2013}]{ztj}
Yu H., Yuan S., Zhang T., PRD 88, 103528.

\bibitem[\protect\citeauthoryear{Zhao et al.}{2012}]{zgb}
Zhao G., et. al., PRL. 109, 171301.

\bibitem[\protect\citeauthoryear{Zhang \& Ma}{2013}]{zsn}
Zhang S., Ma Y., PRL, arXiv:1303.6124.

\end{thebibliography}
\end{document}